# Study on Virtual Gear Hobbing Simulation and Gear Tooth Surface Accuracy

Zhi Geng*‡  , Gang Li**

*Shanghai Macrockets Technology Co., Ltd, Minhang, 201101, Shanghai, China

**Department of Mechanical Engineering, University of Maryland, Baltimore County, Baltimore, MD 21250, USA

(gengzhi_vali@163.com, gangli@umbc.edu)

‡ Corresponding Author; Zhi Geng, 201101, Shanghai, China, Tel: +86 138-1645-4123,

Fax:  +86 0539 5921380, gengzhi_vali@163.com

**Abstract-** This paper presents a digital simulation method for the hobbing process of cylindrical gears. Based on the gear generation principle, taking the professional software as the tool, the problem of virtual hobbing simulation on involute helical gears was studied, and virtual hobbing simulation of hobbing on the whole gear was completed by using macros of CATIA V5. The validity of this method was validated through analyzing the tooth surface accuracy error of the model which was below 1μm between the virtual tooth surface and the theoretical tooth surface and the possible factors affected the tooth surface accuracy during manufacturing was also carried on the discussion. It offers a fictitious three-D platform for studying the principle of gear cutting machine's error as well as the finite element analysis between the ideal tooth surface and the erroneous tooth surface.

**Keywords** gear, gear hobbing, optimization, gear manufacturing, simulation.

## 1. Introduction

With the development of modern industrial technology of gears, the high accuracy gear model becomes urgent need to study the mechanical properties of the gears. Now there are two ways available on the involute gear modeling: direct drawing and simulation process. Due to the complex gear shape, it is more difficult to complete three-dimensional modeling correctly by directly drawing, in particular, the part of the tooth root transition curve [1-3]. Gear is a core mechanical component widely used in automotive, aerospace, high-speed rail, energy equipment and other industries. Gear processing is a complex process [1]. To achieve targets of high-efficiency and high-power of gearboxes, the integration of a high-speed motor and a gearboxes is a development trend of a drivetrain system [2,3]. A new drivetrain scheme of a two-speed planetary gearbox and a one-way clutch can well match with the high-speed motor. A vehicle transmission transmits the rotating power of the energy source, whether an electric motor or an internal combustion engine (ICE), through a set of gears to a differential, the unit that spins the wheels [4-6]. Any vehicle, ICEs or electric vehicles, needs more torque than speed to propel the car from a dead stop, and more speed than torque once the vehicle already has forward momentum [7,8]. Comparing with traditional multi-speed gearboxes for ICEs, the two-speed gearbox is simplified but has a high requirement on its reliability [9]. The reliability performance of the two-speed planetary gearbox of electrical vehicles is closely related to the characteristics of transmission errors of its gear pairs [10].

Cylindrical involute gears, i.e., spur and helical ones, are widely used in gearboxes and planetary gear trains in many other industrial applications [11-13]. Evolution of the design and manufacture of such gears by hobbing, shaping, and grinding has been impressive. Geometry, design and manufacture of helical gears was the subject of research represented in the works [14-16] and many others. Generally, gears are manufactured via hobbing [17,18] or forming cutting [19-21] based on the theory of gearing. For some gears with special tooth profiles, e.g., concave-convex and spiral tooth profiles, their manufacturing methods and machine-tools are complex. Since meshing performances of these gears with special tooth profiles are highly sensitive to manufacturing errors [22,23], high manufacturing accuracy of gear machine-tools is required for these gears [24-27].

Transmission errors and contact patterns are two typical methods for meshing performances evaluation of gear systems [10,28,29]. A tooth profile modelling method was developed to improve accuracy of tooth contact analysis for gear tooth profiles [30]. Some other meshing performances, e.g., power losses, can also be evaluated based tooth contact analysis [31-33]. Since these gears have convex-concave tooth profiles, they cannot be manufactured via standard gear manufacturing methods. During a manufacture in this way, for each of the gear modules and the radius of curvature, a different blade size and gear holder is needed. However, it's clear that these gears have many advantages, if they can be produced sufficiently in the industry [34,35]. Since these gears have better load-bearing capabilities, have a balancing feature for the axial forces, quiet operation feature and their lubrication characteristics is better than herringbone gears and spur gears [36-38]. It's noteworthy that there are number of studies carried out recently in relation to these gears [39,40]. Many applications, e.g., automobile [41,42] and tunnel boring machines [43,44] require high reliability of gearboxes.

Application of modern CNC for gear form grinding methods is introduced new concepts in design and formed of involute gears with modifications. The study describes a new reliability-enhanced form grinding method based on a predesigned second-order transmission error function. The proposed method is based on the kinematical modelling of the basic machine settings and motions of a virtual generating rack cutter. This work focus on the design of gear drives with reduced noise is based on application of a predesigned parabolic function of transmission errors. Taking CATIA V5 as the modeling platform, using the macro function, through simulating the reality gear hobbing method, the whole gear hobbing simulation technology of involute helical gear was achieved, a model of high-precision gear was obtained, and the factors that may affect the accuracy of tooth surface during processing gear was also discussed. This research method could be a reference on gear grinding and other types of processing gear.

In this study, a digital simulation method for the hobbing process of cylindrical gears is proposed. Firstly, a three-dimensional meshing simulation model of hobbing cutters and cylindrical gears is established. Secondly, the relationship between the angle of hobs and gears in hobbing process is analyzed. Thirdly, by solving the intersection point between the tangent length of tooth flanks and the hobbing involute helicoid, the simulated tooth profile of the hobbing process is obtained. The simulation method can improve the gear processing accuracy and gear digital manufacturing.

## 2. Principle of the Hobbing Process

A hobbing process is realized by the gear-generating method, as shown in Fig. 1. In the process, based on the tooth profile meshing principle, the tooth profile is machined. The hobbing cutter and the gear blank are transmitted according to a fixed transmission ratio and the gear profile conjugated with the hob profile can be machined on the billet through the cutting motion of hobbing cutter [18]. At the same time, the whole gear can be machined when the hobbing cutter feeds along the axis of the tooth blank. The hobbing process of the tooth blank is equivalent to the meshing process of worm and gear [18].

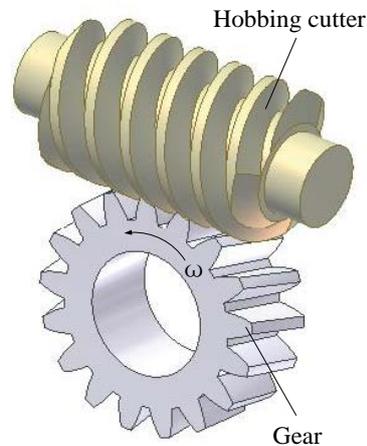

**Fig. 1.** Schematic diagram of the hobbing principle.

## 3. Simulation Method of Helical Gear Hobbing based on CATIA

Gear hobbing is in accordance with the principle of engagement between the gear and the rack. The normal cutaway section of the hob has the rack shape when the hob high-speed rotating around the center axis the tooth of cutter can be regarded as the rack infinitely long.

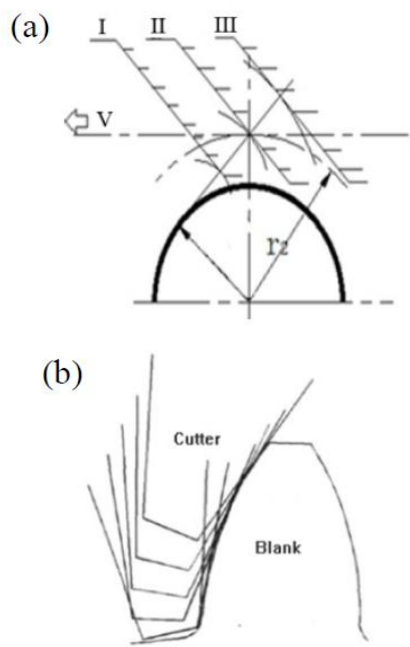

**Fig. 2.** (a) Relative motion of the rack and gear blanks, and (b) the schematic plan of gear cutting.

In Fig.2 (a), the rack goes forward with speed V, the gear rotates with a certain angular velocity, $r_2$ is the gear pitch radius, and to meet that linear velocity of the rack equal to the product of gear angular velocity and the pitch radius, so the pitch line of rack rolls without sliding with gear, and the rack takes the place I、II、III one by one. In this way, because of the rack's moving on the direction perpendicular to the gear axis, the involute curve is given. Fig.2 (b) shows that rack cutter is cutting into each position of the gullet sequentially, every relative moving to gear blank cuts a thin layer of metal from the gear blank. The relative motion of tool and gear blank can be achieved by the command of translation and rotation in CATIA V5 [5]

The gear hob is continuous from the visual point of view when it is in high-speed rotation, similar to the trapezoidal thread. The actual gear hob teeth is not continuous, which is beneficial to cooling and chip removal, through its high-speed rotation can form a continuous cutting edge. In cutting simulation, cutting process is different from the actual one. That is, to achieve the gear hob cutting through the CATIA internal subtract of the Boolean operation. Therefore, the cutter can be simplified to be consistent with the actual processing of the continuous cutting edge in the simulation, as shown in Fig. 3. The detail geometry of gear hob modeling is shown in Table 1. The gear blank model is a cylinder, which is relatively simple to build. In CATIA, it can be developed by stretch command or rotation command.

**Table 1.** Detail geometry of gear hob modeling

| Item | Value |
|---|---|
| Module | 2 |
| External Diameter | 71 |
| Aperture | 27 |
| Gear width | 50 |
| Pitch Diameter | 65.06 |
| Addendum | 2.5 |
| Whole Depth | 5 |
| Normal Pitch | 6.286 |
| Normal Tooth Thickness | 3.143 |
| Axial Profile Angle | 20°1′ |
| Spiral Lead Angle | 1°46′ |
| Hob Thread Direction | Right |
| Tip Radius | 0.6 |
| Fillet Radius | 0.6 |

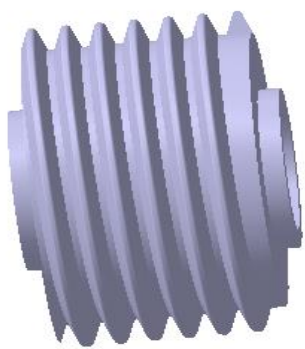

**Fig. 3.** Outline drawing of hob rotation.

The gear hob is continuous from the visual point of view when it is in high-speed rotation, similar to the trapezoidal thread. The actual gear hob teeth is not continuous, which is beneficial to cooling and chip removal, through its high-speed rotation can form a continuous cutting edge. In cutting simulation, cutting process is different from the actual one. That is, to achieve the gear hob cutting through the CATIA internal subtract of the Boolean operation. Therefore, the cutter can be simplified to be consistent with the actual processing of the continuous cutting edge in the simulation, as shown in Fig. 3. The detail geometry of gear hob modeling is shown in Table 1. The gear blank model is a cylinder, which is relatively simple to build. In CATIA, it can be developed by stretch command or rotation command.

This simulation flow chart of the gear hobbing process is shown in Fig. 4. After starting CATIA program, we first should build the modeling of gear hob and gear blank, then adjust the installation angle of the gear hob according to the principle that right-hand hob machining right-hand gear and finally transfer the hob and gear blank to the location of the initial processing. After that we run the VBA macro, which drives CATIA adjust the relative position of cutter and gear blank, and then does a subtract of Boolean operation between the hob and gear blank, removing the position that occupied by cutter on the gear blank, and determines whether the simulation finish a whole tooth width, If not finish the cut, then repeats the above steps to adjust the tool to the next position, if finished the whole tooth width, the cutting process is done; the complete hobbing process is shown in Fig. 5. Figure 5(a) shows that the hob start cutting the gear blank from the initial position, with the hob feed amount increasing, the hob cutter is moving along the axial direction, as shown in Fig. 5(b), the gear cutting continues, when the hob cutter feeds the whole tooth width, the gear cutting ends, as shown in Fig. 5(c), at this time, a complete helical model is obtained, as shown in Fig. 5(d).

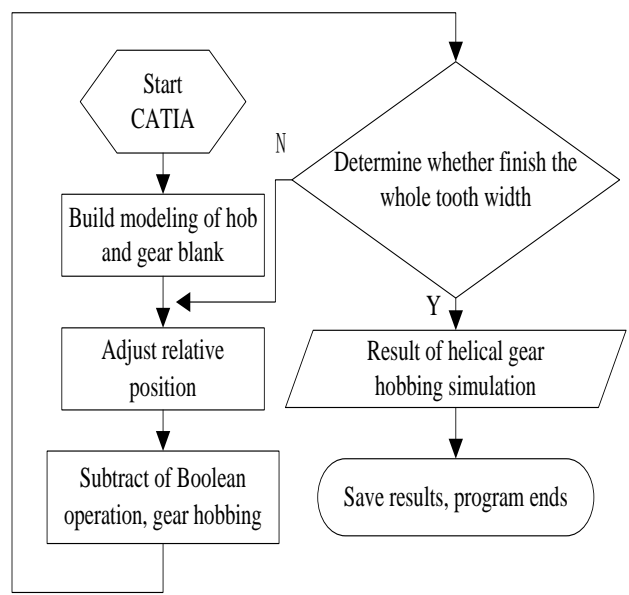

**Fig. 4.** Flowchart of gear hobbing simulation process using CATIA.

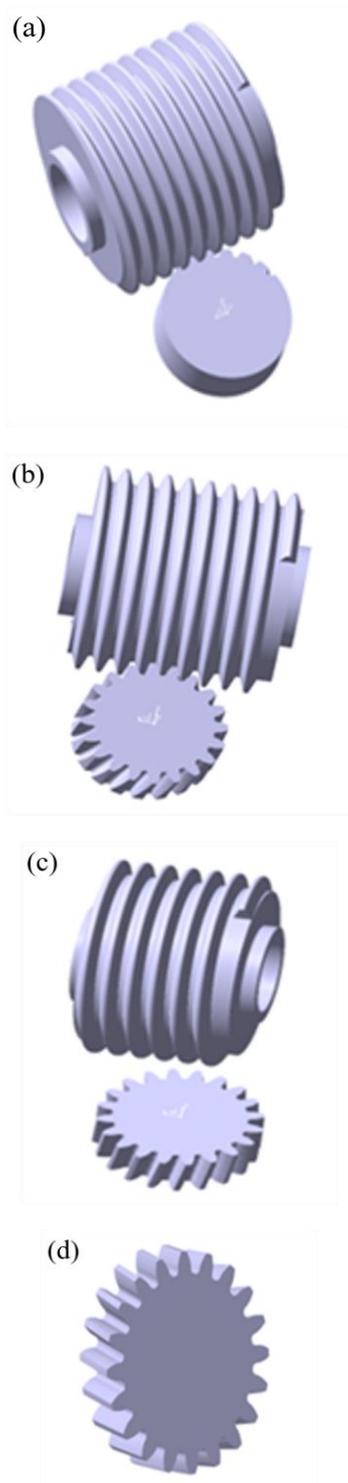

**Fig. 5.** (a) The beginning of cutting, (b) in the cutting, (c) cutting completed, and (d) simulation processed gear.

## 4. Discussion of Tooth Surface Accuracy

### 4.1. Method of Helical Virtual Tooth Surface Modelling Setup

Cutting process of virtual model machining is achieved by subtract of Boolean operation, VBA macro controls hob and gear blank moving in accordance with some regular rules. It will remove the position occupied by the hob from the gear blank each time, when the hob is adjusted to the cutting position, repeat what was said above, the helical gear hobbing will be finished.

During the process of gear generating, the tool cutting edge cuts out intersection of two adjacent surfaces on the gear blank, known as the tool mark lines [5]. It will leave a lot of tool mark lines when the tool is moving, as shown in Fig. 6.

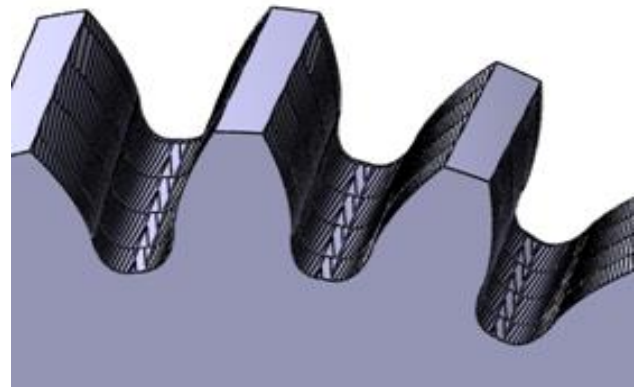

**Fig. 6.** Tooth surface with tool mark lines.

Until now, the hobbed tooth surface is connected by a series of small pieces, these pieces is obtained by each subtract of Boolean operations that removing the position occupied by the hob from the gear blank, is surrounded by the adjacent tool mark lines. It also can extract three-dimensional coordinate data of the hobbed gear surface. However, the mesh of small surfaces will make adverse effects, if it is run finite element analysis directly. It is described that the reconstruction method of tooth profile surface in [5], but it is little trivial. In this paper, virtual tooth surface, as shown in Fig. 7, is obtained by extracting and combining these pieces of small surfaces together through CATIA commands.

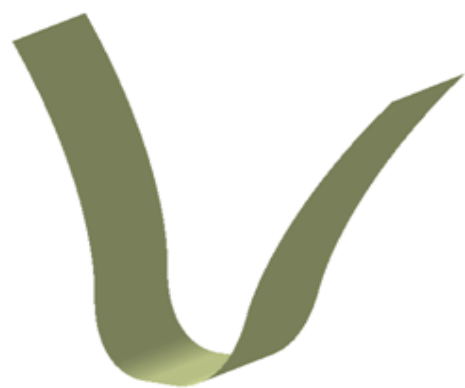

**Fig. 7.** Virtual gear tooth surface.

### 4.2. Authentication Method of Virtual Helical Gear Tooth Surface Accuracy

In order to study the accuracy of simulated tooth surface, we first introduce the method of acquiring tooth surface error data. The theoretical tooth surface data can be calculated by the theory of involute helicoids, and can be imported to CATIA to finish accuracy verification. There are 25 discrete points, 5 rows and 5 columns, which are selected equidistantly from theoretical calculations of tooth surface point cloud 25 x 25 (plane projection is shown in Fig. 8). Adjustment steps for tooth surface error data are:

① Import the theoretical point to CATIA, make sure that the theoretical gear axis coincides with the axis of model, and then the theoretical points cloud can be rotated any angle around the gear axis among 0-360°.

② Adjust the angle that the theoretical points in the circumference to make sure the center point of the 25 points (row 3, column 3) coincides with the virtual tooth surface, as shown in Fig. 9, the distance is 0 mm.

③ Then take the normal distance (shortest distance) as a tooth surface error between theory points and virtual tooth surface one by one.

Based on the steps above, the error data of virtual tooth surface can be got.

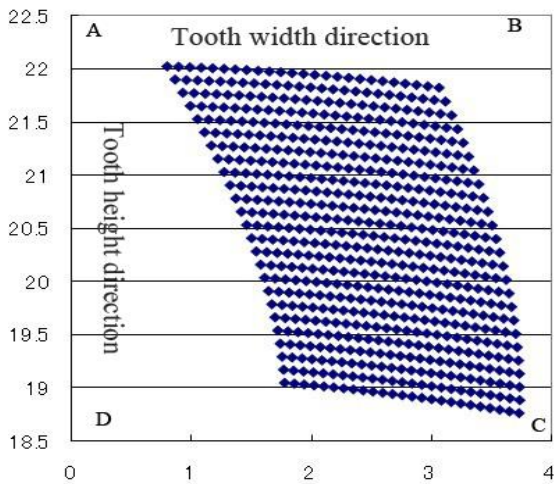

**Fig. 7.** Flat projection of theoretical gear tooth surface points cloud 25 x 25

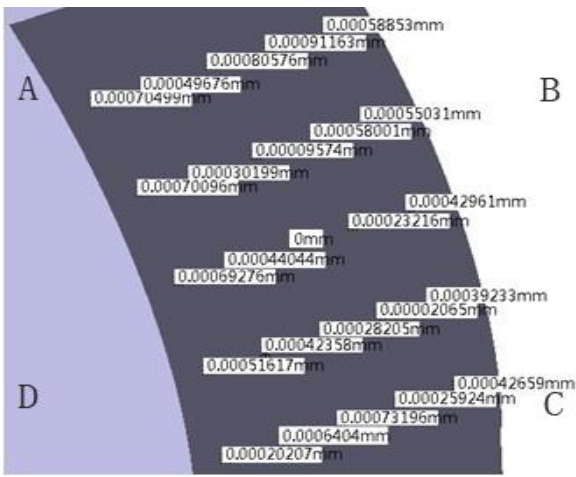

**Fig. 8.** The distance between center point and virtual gear tooth surface is 0 mm

*4.3. Accuracy Verification of the Virtual Tooth Surface and Its Influencing Factors*

To set up experiments to determine the influence trend of factors based on the assumed factors: hob feed speed (amount of feed after each loop) (mm/r) and interval angle of gear blank rotation (°). Taking the maximum error data as the research object, experimental results are shown in Table 2.

From Analysis of Table 2 we can see, when interval angle of the gear blank rotation is certain, and hob feed reduces among some extent, the maximum error of tooth surface will also reduce. When the hob feed is certain, the reducing of interval angle of gear blank rotation has a significant influence to the accuracy of the tooth surface, which is consistent with the reality. In actual processing, the gear blank is continuous rotation, the simulation machining cannot be continuous movement, so we disperse the continuous movement to a finite number of rotation angle, the smaller the rotation angle is, the closer to the actual continuous movement, at this time, the resulting virtual tooth surface is more smooth, so it has a higher accuracy when compared the simulation accuracy between a large rotation angle and a small rotation angle. Due to computer hardware (memory) constraints, it will take a particularly long time to calculate when selects a small interval angle, and finally it will be out of calculation because of lack of memory, so we just take 2 as the smallest interval angle, and while the hob feed is generally not less than 1, otherwise it becomes too slow, leads to a longer processing cycle , and now it has a very high accuracy, with the development of computer hardware, more accurate simulation can be achieved according to the analysis of the trends in the paper.

Using the methods described above extracts 25 error data to form the error curved surface, as shown in Fig. 10, the maximal error is 0.91163 μm. The position of ABCD shown in Fig. 10 coincides with Figs. 8 and 9 correspondingly. It can visually see that distribution of error on the virtual tooth surface through the error curved surface. Further research on the surface error forming principle of the gear hobing machine and its compensation method can be studied by using the method explained above.

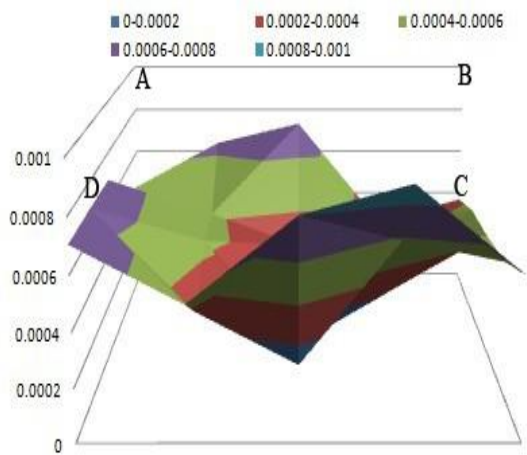

**Fig. 10.** The error curved surface (maximum error is 0.91163 μm)

**Table 2.** Results of influencing factors (maximum error of tooth surface)

| Hob feed speed (mm/r) | Interval angle of gear blank rotation (°) | | |
|---|---|---|---|
| | 8 | 4 | 2 |
| 5 | 11.20932 | 8.66371 | 7.41972 |
| 4 | 5.72456 | 3.20556 | 3.53586 |
| 3 | 3.16018 | 2.42126 | 1.52903 |
| 2 | 2.75746 | 1.40773 | 1.11871 |
| 1 | 2.27441 | 1.15853 | 0.91163 |

## 5. Conclusion

Based on the gear generation principle, taking the professional software as the tool, the problem of virtual hobbing simulation on involute helical gears was studied, and virtual hobbing simulation of hobbing on the whole gear was completed by using macros of CATIA V5. This method has the following characteristics:

1. The accurate gear tooth surface was obtained by extracting and integrating tiny curved faces which were made through virtual hobbing. The validity of this method was validated through analyzing the tooth turface accuracy error of the model which was below 1μm between the virtual tooth surface and the theoretical tooth surface, all above is easy to achieve.

2. The method of virtual hobbing simulation on involute helical gears was accomplished. It offers a fictitious three-D platform for studying the principle of gear cutting machine's error as well as the finite element analysis between the ideal tooth surface and the erroneous tooth surface.

The method also provides an effective technical support for the simulation of other methods, such as virtual simulation of the gear-grinding machining.

## Acknowledgements

This work was financially supported by Projects from National Natural Science Foundation of China (51075279), and Innovation Program of Shanghai Municipal Education Commission (10ZZ92).